\documentclass{webofc}
\usepackage[varg]{txfonts}   
\usepackage{hyperref}
\usepackage{url}
\usepackage{lineno}

\newcommand\pT{\ensuremath{p_\mathrm{T}}}

\hypersetup{colorlinks=true,citecolor=blue,urlcolor=blue,linkcolor=blue}

\begin{document}
\title{Probing hadronization with the charge correlator ratio in $p$+$p$, $Ru$+$Ru$ and $Zr$+$Zr$ collisions at STAR}
\author{Youqi Song\inst{1}\fnsep\thanks{\email{youqi.song@yale.edu}} for the STAR Collaboration}
\institute{Wright Laboratory, Yale University}
\abstract{The parton-to-hadron transition, known as hadronization, is dominated by non-perturbative Quantum Chromodynamics (QCD) effects and thus challenging to study from first-principle calculations. On the other hand, experimental studies of observables sensitive to hadronization could provide valuable input. The charge correlator ratio $r_c$ studies the charge correlation between the leading two hadrons in jets and is sensitive to hadronization effects. With data taken at $\sqrt{s_{\rm{NN}}}=200$ GeV at STAR, we measure $r_c$ in $p$+$p$ collisions to probe for string-like fragmentation, and in $Ru$+$Ru$ and $Zr$+$Zr$ collisions to probe for potential modification to hadronization in the Quark-Gluon Plasma (QGP). These measurements, compared with various model predictions, are expected to distinguish between phenomenological model descriptions for hadronization in vacuum, and provide insight into jet-medium interaction in the QGP.}

\maketitle
\section{Introduction}
\label{intro}
In high-energy particle collisions, jets can be created in hard scattering, and encode information from the subsequent processes of the parton shower and hadronization. Since hadronization is dominated by non-perturbative effects, it is challenging to understand from first-principle QCD calculations, thereby highlighting the crucial role of experimental inputs. Investigation of the internal structure of jets, known as jet substructure, can shed light on the dynamics of jet evolution.

One novel jet substructure observable, the charge correlator ratio $r_c$, is sensitive to hadronization effects \cite{chien2022probing}, \cite{h1}. For leading and subleading hadrons in jets $h_1$ and $h_2$, $r_c$ quantifies the tendency of production of $h_2$ or its anti-particle $\bar{h_2}$, in the presence of $h_1$ production. It is defined as 
\begin{equation}
    r_c(x)=\frac{d\sigma_{h_1h_2}/dx-d\sigma_{h_1\bar{h_2}}/dx}{d\sigma_{h_1h_2}/dx+d\sigma_{h_1\bar{h_2}}/dx},
\end{equation}
where $x$ is a generic kinematic variable such as the jet \pT. Specifically, for inclusive charged hadrons, $h_1h_2$ denotes that the leading and subleading hadrons in jets carry the same electric charge, while $h_1\bar{h_2}$ denotes that they carry opposite electric charges. In the extreme limit where string-like fragmentation is the only mechanism for hadron production, we expect a perfect charge correlation between a hadron and an anti-hadron, $r_c\rightarrow-1$; in the other limit where the environment for hadron production is an infinite charge bath with no net charge, we expect no charge correlation between pairs, $r_c\rightarrow0$. We anticipate the measurement of $r_c$ to be between $-1$ and $0$, with its exact value sensitive to the fragmentation mechanism.

Jets produced in heavy ion collisions interact with the soft partons in the QGP and lose energy. Uncovering the potential modification of jet substructure due to the presence of the QGP is important for understanding of jet-medium interaction. By measuring $r_c$ in heavy ion collisions, we probe for potential modification to hadronization in the core of the jet \cite{chien2023probing}.

In these proceedings, we present measurements of the charge correlator ratio $r_c$ in jets in $p$+$p$ collisions at $\sqrt{s}=200$ GeV at STAR, to probe for string-like fragmentation. We compare the results with predictions from Monte Carlo event generators. In addition, we present ongoing studies on the first measurement of $r_c$ in heavy-ion collisions, in $Ru$+$Ru$ and $Zr$+$Zr$ collisions at $\sqrt{s_{\rm{NN}}}=200$ GeV at STAR. 

\section{\texorpdfstring{$r_c$ }\ in \texorpdfstring{$p$+$p$ }\ collisions}
\label{pp}
\subsection{Analysis details}
\label{pp_analysis}
The STAR experiment \cite{STAR:2002eio} recorded data from $\sqrt{s} = 200$ GeV $p$+$p$ collisions during the 2012 RHIC run. Tracks are reconstructed from the Time Projection Chamber (TPC), and neutral energy deposits are measured from the Barrel Electro-Magnetic Calorimeter (BEMC) towers. Events are required to have primary vertices within $\pm 30$ cm from the center of the detector along the beam axis, and to pass the jet patch trigger which requires a minimum transverse energy $E_{\mathrm{T}}>7.3$ GeV deposited in a $1 \times 1$ patch in $\eta \times \phi$ in the BEMC\footnote{Each patch contains 400 towers.}. We reconstruct jets from TPC tracks ($0.2 < \pT < 30\ \mathrm{GeV}/c$) and BEMC towers ($0.2 < E_{\mathrm{T}} < 30\ \mathrm{GeV}$) using the anti-$k_{\mathrm{T}}$ sequential recombination clustering algorithm \cite{cacciari2008anti} with a resolution parameter of $R=0.4$. We apply the selections of $p_{\mathrm{T}} > 15$ GeV$/c$, $|\eta|<0.6$, transverse energy fraction of all neutral components $<0.9$, and number of charged constituents $N_{ch}\geq2$ on reconstructed jets.

Ref. \cite{chien2022probing} proposes to only include jets whose leading and subleading particles are charged for the measurement of $r_c$. That is, if the leading and/or subleading particle of a truth-level jet is a $\pi^0$, for example, then the jet should not be included for the measurement of $r_c$. However, since each of the decay products of the $\pi^0$ (most likely $\gamma$) measured shares only a fraction of the parent \pT, at the detector level, the jet could still be misidentified as having leading and subleading charged particles. Due to this complication, we use a definition slightly different from that in \cite{chien2022probing} for our measurement and Monte Carlo simulations; we measure the charge correlation between the leading and subleading track pairs in jets, regardless of whether there is any neutral constituent with a higher energy.

We fully correct for detector effects through a two-step procedure. First, we carry out ``mistagged subtraction'' on $r_c$ to account for possible misidentification of tracks that are not leading or subleading, due to tracking inefficiency. Then, we apply a bin-by-bin jet \pT\ correction to account for jet energy scale.

\subsection{Results}
\label{pp_result}
\begin{figure}[ht]
    \centering
    \includegraphics[width=5cm,clip]{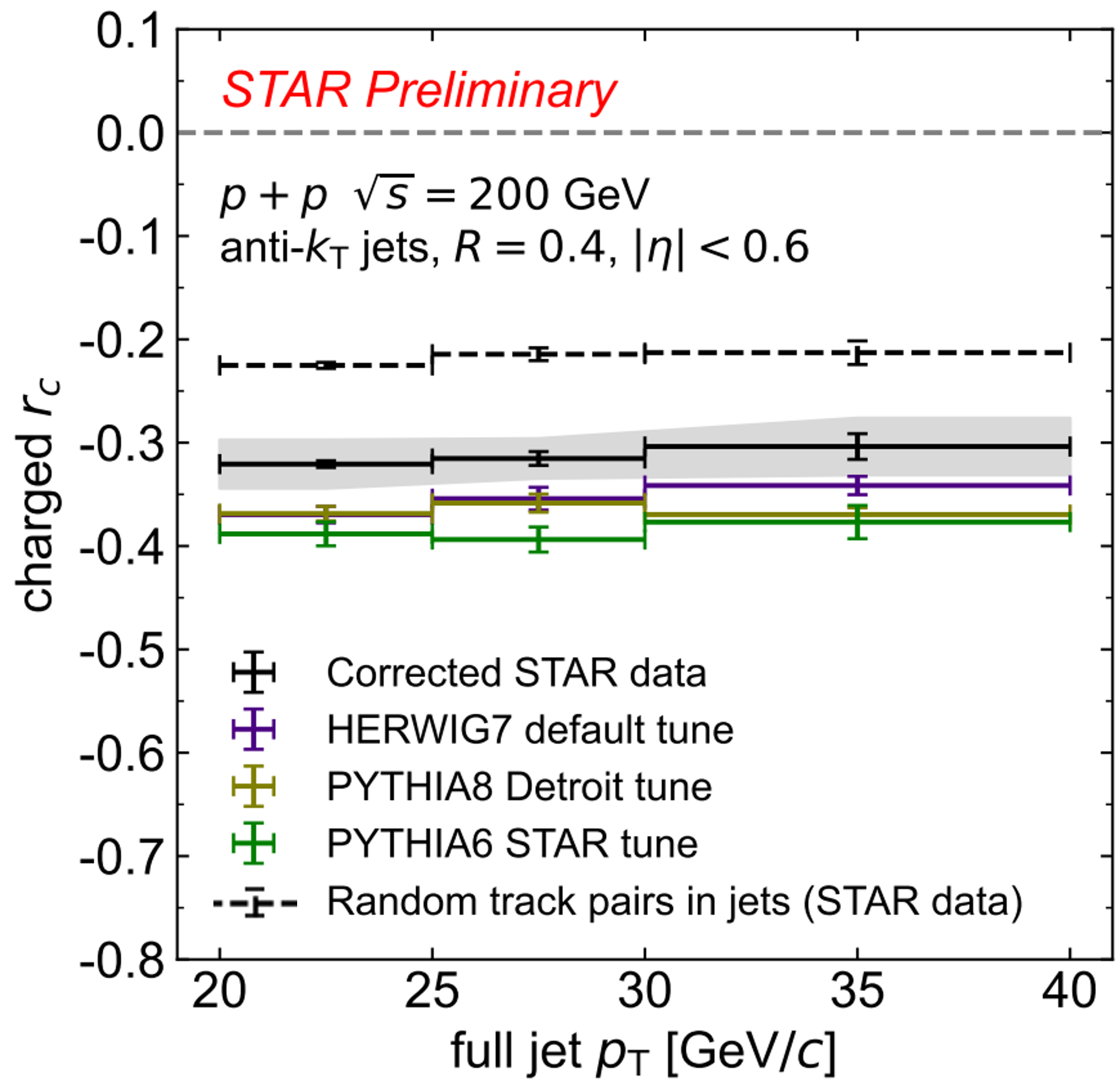}
    \caption{$r_c$ as a function of jet \pT\ in $p$+$p$ collisions, for the fully corrected STAR data (black solid), with comparison with predictions from HERWIG7 \cite{herwig} (purple), PYTHIA8 \cite{pythia8} Detroit tune \cite{aguilar2022pythia} (gold) and PYTHIA6 \cite{Sjostrand:2006za} STAR tune \cite{Adkins:2015ccl} (green). Effective $r_c$ for random track pairs in jets in STAR data is shown in black dashed lines.}
    \label{pp_rc}
\end{figure}

Figure~\ref{pp_rc} shows the $r_c$ as a function of jet \pT, for the fully corrected STAR data in black solid lines, with comparison with predictions from HERWIG7 \cite{herwig}, PYTHIA8 \cite{pythia8} Detroit tune \cite{aguilar2022pythia} and PYTHIA6 \cite{Sjostrand:2006za} STAR tune \cite{Adkins:2015ccl}, in purple, gold and green lines, respectively. As expected, the $r_c$ values lie below $0$, which is the limit of a charge bath. 

Figure~\ref{pp_rc} also shows in black dashed lines the effective $r_c$ as a function of jet \pT\ for random track pairs in jets in data. We randomly sample two tracks (without replacement) from a jet, and their charge information is used for calculation of the effective $r_c$. The effective $r_c$ values for the random tracks are negative (around $-0.2$). This shows that there is already some effect of local charge conservation between the random tracks, and is consistent with the result that the average and peak values of jet charge are around $0$ for jets measured in similar kinematics at STAR \cite{jet_charge}.

Compared with the effective $r_c$ for random track pairs, the $r_c$ values are more negative (around $-0.3$) with the requirement for the tracks to be leading and subleading in jets, highlighting the additional correlation from fragmentation. The $r_c$ values from both the STAR data and event generator predictions exhibit no jet \pT\ dependence in $20 < \pT < 40$ GeV$/c$. Given that PYTHIA uses the Lund string fragmentation model while HERWIG uses the cluster hadronization model, it is interesting that both event generators underpredict $r_c$ in data to a similar extent.

We further investigate if effects other than fragmentation could influence the value of $r_c$. Using events generated with HERWIG7 and PYTHIA8 (default tune), we cluster jets with the anti-$k_{\rm{T}}$ algorithm and a radius of $R=0.4$, with selections of $\pT > 20$ GeV$/c$ and $|\eta| < 0.6$, and study the origin of $\pi^+$ if they are the leading tracks in jets. We find that about $50\%$ of the $\pi^+$ come from quarks or diquarks in PYTHIA, while only about $30\%$ of them come from clusters in HERWIG. Since in the event generator simulations we have disabled hadronic decays mediated by the electroweak forces, the rest of leading $\pi^+$ in jets come from resonance decays mediated by the strong force. The effect of resonance decays reduces $r_c$'s sensitivity to fragmentation, although in sign-preserving decays such as $\rho^+\rightarrow\pi^+\pi^0$, the $\pi^+$ maintains the charge information from its parent $\rho^+$. It is possible that this discrepancy between HERWIG and PYTHIA in fragmentation vs. resonance decay fractions arises from their different hadronization mechanisms, so additional measurements of resonance production in $p$+$p$ collisions might also help distinguish hadronization models.

\section{\texorpdfstring{$r_c$ }\ in \texorpdfstring{$Ru$+$Ru$ }\ and \texorpdfstring{$Zr$+$Zr$ }\ collisions}
\label{heavy_ion}
The STAR experiment recorded data from $\sqrt{s_{\rm{NN}}} = 200$ GeV $Ru$+$Ru$ and $Zr$+$Zr$ (isobar) collisions during the 2018 RHIC run. Tracks and neutral energy deposits are measured from the TPC and BEMC, respectively. Events are required to have primary vertices within $(-35,25)$ cm along the beam axis from the center of the detector, and within 2 cm radially away from the beam axis. In addition, to reject pileup events, the location of the primary vertex along the beam axis reconstructed from the TPC is required to be within 5 cm from that reconstructed from the Vertex Position Detector. The charged particle multiplicity in $|\eta| < 0.5$ measured from the TPC is used for centrality determination, as detailed in \cite{cme}.

\begin{figure}[ht]
    \centering
    \includegraphics[width=0.5\linewidth]{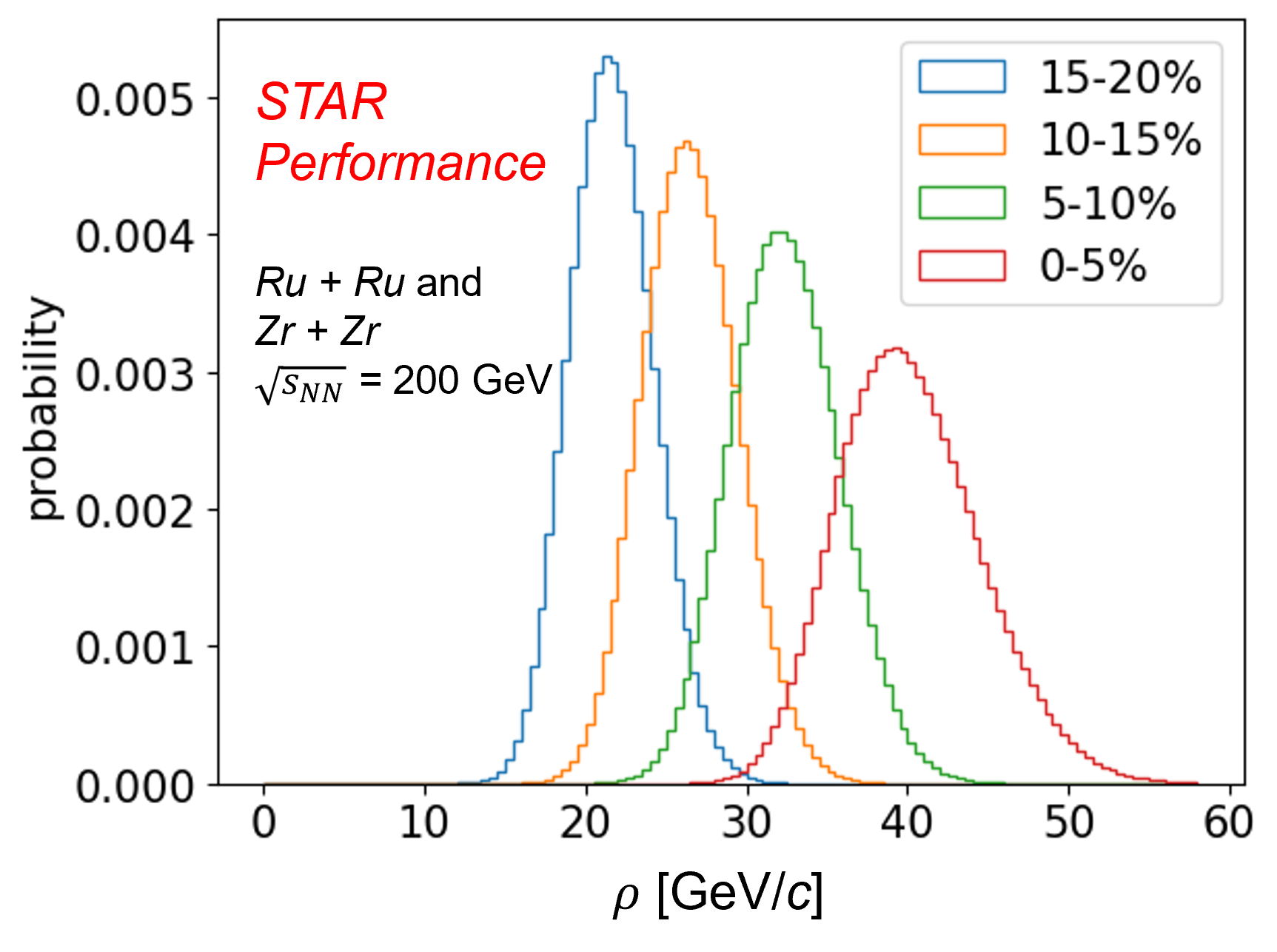}
    \caption{$\rho$ distribution for isobar collision events, with centralities of $0-5\%$ (red), $5-10\%$ (green), $10-15\%$ (orange), and $15-20\%$ (blue).}
    \label{fig:rho}
\end{figure}

To understand the background present in isobar collision events, we also reconstruct jets using the $k_{\rm{T}}$ \cite{kt} clustering algorithm with $R=0.4$ and $|\eta|<0.6$. After excluding the hardest two jets in each event, we calculate the background momentum density $\rho=\mathrm{median}(p_{\mathrm{T}i}/A_i)$ with the remaining jets, where $A_i$ denotes the jet area of jet $i$. Figure \ref{fig:rho} shows the $\rho$ distribution for the four most central centralities, $0-5\%$ in red, $5-10\%$ in green, $10-15\%$ in orange, and $15-20\%$ in blue. Studies have shown that combinatorial jets can be suppressed with a background-subtracted \pT\ selection of $\pT-\rho A>20$ GeV$/c$ \cite{He:2024roj}.

Next, to understand how the two leading tracks in jets are affected by the background, we create a toy sample by embedding PYTHIA jets into isobar data. We generate events of $p$+$p$ collisions at $\sqrt{s}=200$ GeV with PYTHIA8 Detroit tune, and for each event with at least one jet that passes the jet selection in Section \ref{pp_analysis}, we embed all the jet constituents into an event from $0-20\%$ central isobar collisions. Then with the embedded events, we re-cluster jets using the anti-$k_{\rm{T}}$ clustering algorithm with $R=0.4$ and $|\eta|<0.6$, with the same track and tower selection as detailed in Section \ref{pp_analysis} and jet $\pT-\rho A>20$ GeV$/c$, where $\rho$ is calculated with the embedded events as well. Among all these jets, we find that $21\%$ of the jets (1) contain a leading or subleading track from the isobar data; and (2) are not  ``combinatorial''\footnote{The ``combinatorial'' jets in the toy sample may be real jets from the central isobar collisions, but we call them ``combinatorial'' in the sense that they are not produced in PYTHIA, so they are not ``signal'' in the toy model. They are identified with an axis more than $0.4$ away from the corresponding PYTHIA jet axis.}. This significant fraction suggests that even though the leading and subleading tracks in jets are more robust against background contamination than inclusive hadrons \cite{chien2023probing}, to measure $r_c$ in central isobar collisions, we still need to account for jets arising from fragmentation while containing background particles.

To account for all sources of background contamination, we use
\begin{equation}
    \label{eq:rc_raw}
    r_c(\mathrm{raw\ data})=P(\mathrm{comb})\cdot r_c(\mathrm{comb}) + P(\mathrm{BB})\cdot r_c(\mathrm{BB}) + P(\mathrm{SB})\cdot r_c(\mathrm{SB}) + P(\mathrm{SS})\cdot r_c(\mathrm{SS}),
\end{equation}
where $P$ denotes probability, ``comb'' is short for combinatorial jet, ``BB'' stands for ``Background-Background'', meaning that the jet is not combinatorial but both leading tracks in jets are background particles, ``SB'' stands for ``Signal-Background'', meaning that one of the leading tracks in jets is a background particle, and ``SS'' stands for ``Signal-Signal'', meaning that both leading tracks in jets are from fragmentation.

\begin{figure}[ht]
    \centering
    \includegraphics[width=0.35\linewidth]{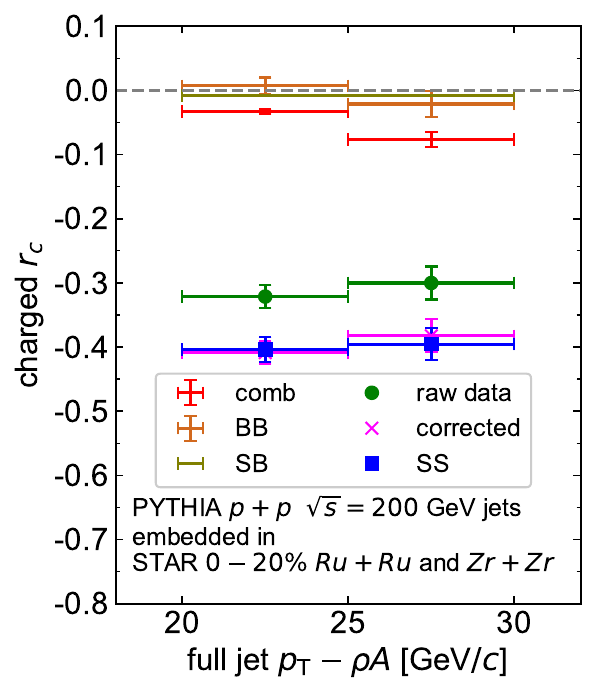}
    \caption{$r_c$ from various contributions for two selections of background subtracted jet \pT. }
    \label{fig:closure_aa}
\end{figure}

We verify Equation \ref{eq:rc_raw} with the toy embedding sample, treating particles from PYTHIA jets as signal and particles from the isobar data as background. Figure \ref{fig:closure_aa} shows the values of $r_c$ from various contributions. We determine $r_c(\mathrm{raw\ data})$\footnote{$r_c(\mathrm{raw\ data})$ is estimated using the embedding only for this toy study, and will be measured with the actual data in the future.}, $r_c(\mathrm{BB})$\footnote{To estimate $r_c(\mathrm{BB})$, we consider all jets that pass our selections from the embedding, and find the $r_c$ using the leading and subleading background track charge information, even if the jet contains a signal track with a higher \pT.} and $P$ factors from the embedding, and estimate in a data-driven way with the $0-20\%$ central events $r_c(\mathrm{comb})$ and $r_c(\mathrm{SB})$\footnote{We estimate $r_c(\mathrm{SB})$ using data instead of the embedding due to limited statistics. To estimate $r_c(\mathrm{SB})$, we cluster jets without background subtraction and exclude the two leading jets. We find $r_c(\mathrm{SB})$ to be independent of jet \pT. With jets from embedding, we have confirmed that $r_c(\mathrm{BB}) \approx r_c(\mathrm{SB})$, although with large uncertainties.}. Then we calculate $r_c(\mathrm{SS})$ using Equation \ref{eq:rc_raw}. The $r_c(\mathrm{SS})$ values obtained this way, shown in pink crosses, agree with those obtained directly using the PYTHIA information from the embedding, shown in blue squares. This agreement demonstrates that we achieve closure with the background subtraction technique explained above.

On the other hand, when we move to data analysis with the actual isobar data, while we can use the decomposition of different sources of background as given by Equation \ref{eq:rc_raw}, to find the $P$ and $r_c$ factors for each term, we need to use the mixed-event technique and background correlation estimation with perpendicular cones. For example, mixed events, constructed by properly sampling particles from different events to ensure no physical correlation, can be used to estimate $r_c(\mathrm{comb})$; perpendicular cones, which are clusters of particles about $\Delta \phi=\pi/2$ away from jets, can be used to estimate $r_c(\mathrm{BB})$.

\section{Conclusions}
\label{conclusion}
In summary, to probe for string-like fragmentation, we measure $r_c$ as a function of jet \pT\ in $\sqrt{s}=200$ GeV $p$+$p$ collisions at STAR. Compared with predictions from event generators, the fully corrected data show a weaker correlation between the leading and subleading tracks in jets. The significant difference between HERWIG and PYTHIA in their relative fractions of fragmentation vs. resonance decays might explain why their predictions for $r_c$ are similar despite different approaches to fragmentation.

In addition, we present progress towards the first measurement of $r_c$ in heavy ion collisions, to probe for potential modification to hadronization due to the presence of the QGP. With data recorded by STAR from $\sqrt{s_{\rm{NN}}}=200$ GeV $Ru$+$Ru$ and $Zr$+$Zr$ collisions, we study jets produced in $0-20\%$ centrality events, discuss sources of background contributions, and demonstrate closure with our background subtraction technique.


\end{document}